\documentclass{fromcimento}

\usepackage{graphicx}  
\usepackage{amssymb}
\title{The Super$B$ project}
\author{M.~Rama\from{ins:x}}
\instlist{\inst{ins:x} INFN, Laboratori Nazionali di Frascati, Via E.~Fermi 40, I-00044 Frascati, Italy}
\begin{document}

\maketitle

\begin{abstract}
Super$B$ is a next generation asymmetric $e^+e^-$ flavor factory with a baseline luminosity of $10^{36}$~cm$^{-2}$s$^{-1}$, 50-100 times the peak luminosity of the existing $B$-factories. The physics motivation is presented and the complementarity with the LHC is discussed. The conceptual design of the detector is also briefly described. 
\end{abstract}

\section{The role of a super flavor factory in the LHC era}
Super$B$ is a next generation asymmetric $e^+e^-$ flavor factory with very high peak luminosity ($\mathcal{L}=10^{36}{\rm cm}^{-2}{\rm s}^{-1}$) proposed to be built in the Rome area. The main purpose of the experiment is to search for evidence of physics beyond the Standard Model (SM) and investigate its nature. 
 
 The search for new physics (NP) is the main goal of elementary particle physics in the next decade. The search is encouraged by the expectation that the NP mass scale be around 1 TeV, thus directly accessible to the Large Hadron Collider (LHC) at CERN. In this context, it is important to clarify the role of a Super$B$ factory by considering two scenarios depending on whether or not evidence of NP will be found at the LHC.

If direct evidence of NP is found at the LHC, Super$B$ will help determining the flavor structure of NP by constraining its couplings, mixing angles and masses through the measurement of rare decays whose amplitudes are mediated by loops in the SM, or through the observation of lepton flavor violating processes.
The flavor physics observables measured at a Super Flavor Factory provide a set of independent constraints complementary to those measured at high $p_T$ processes. To reconstruct the NP Lagrangian both contributions are required.

If instead the LHC does not discover NP particles, Super$B$ will allow to explore mass scales up to 10 TeV or beyond (depending on the models) and could provide the first evidence of NP.

One may ask how the Super$B$ program compares with the flavor physics potential of LHCb. The outcome of the comparison is that even in this case the complementarity is large.
For example, rare decay modes with one or more neutrinos in the final state such as $B^+\rightarrow l^+\nu$ and $B^+\rightarrow K^{(*)+}\nu\bar{\nu}$, inclusive analyses of processes such as $b\rightarrow s\gamma$ and $b\rightarrow s l^+l^-$ or measurements of the CKM matrix elements $|V_{ub}|$ and $|V_{cb}|$ are unique to Super$B$, where the environment of the $e^+e^-$ collider is clean and relatively simple compared to the events at the hadronic machine~\cite{Browder:2008em}. Some examples will be discussed in sect.~\ref{sec:physics_case}. On the other hand, LHCb has access to measurements that are not possible at Super$B$, such as the time-dependent study of the oscillations of $B_s$ mesons.

It is important to stress that Super$B$ will also provide huge samples of charm hadron and $\tau$ lepton pairs ($2.0\times 10^{10}$ and $1.4\times 10^{10}$ per year, respectively, at the nominal luminosity), enabling powerful studies of NP effects in the up-type quark and lepton sectors with unprecedented precision. Moreover, the machine is being designed to run in a wide range of center-of-mass (CM) energies, down to the $\tau$ threshold and up to the $\Upsilon(5S)$ mass, thus enriching further the physics program.

To reach the goals of Super$B$ in terms of sensitivity, a data sample of about two orders of magnitude larger than the ones accumulated by the current $B$-factories BaBar (0.53ab$^{-1}$) and Belle (0.95ab$^{-1}$) is needed. A collider baseline luminosity of $10^{36}$~cm$^{-2}$s$^{-1}$ would allow to collect 15ab$^{-1}$ per year (1 yr=$1.5\times 10^7$s) corresponding to a data sample of more than 80 billion $B\bar{B}$ pairs in five years of running at the $\Upsilon(4S)$ CM energy.

\section{The physics case of Super$B$}\label{sec:physics_case}
In this section a short selection of measurements that are part of the Super$B$ physics program are briefly discussed. The reader is referred to refs.~\cite{cdr,valencia} for an extensive discussion of the physics reach of the experiment.

\subsection{Rare $B$ decays}
One of the strengths of the physics program of Super$B$ is the large number of decays where the SM uncertainty is small and that can display a measurable deviation from the SM in one or more NP scenarios. A large fraction of these 'golden' channels are rare $B$ decays where NP particles enter the leading loops. A selection of golden modes in different NP scenarios is reported in Table~\ref{tab:golden_modes}. The list of observables in the table is not complete, as well as the number of NP scenarios considered.
\begin{table}[!h]
 \caption{
    \label{tab:golden_modes}
     Golden modes in different New Physics scenarios. An ``X'' indicates the golden channel of a given scenario.
     An ``O'' marks modes which are not the ``golden'' one of a given scenario but can still display a measurable deviation from the Standard Model.
     The label $CKM$ denotes golden modes which require the high-precision determination of the CKM parameters achievable at Super$B$. Table from~\cite{valencia}. }
\begin{tabular}{lccccc}
\hline
   & $H^+$ &~Minimal~&~Non-Minimal~&NP& ~Right-Handed~ \\
   & high tan$\beta$ &FV&FV &~Z-penguins~&currents\\
    \hline
$\mathcal{B}(B \to X_s \gamma)$          &        & X &       O &  & O \\
$A_{CP}(B \to X_s \gamma)$               &         &   &      X  & &O\\
$\mathcal{B}(B \to \tau\nu)$             &X-$CKM$ &   &         & & \\
$\mathcal{B}(B \to X_s l^+l^-)$          &        &   &       O  &O&O \\
$\mathcal{B}(B \to K \nu \overline{\nu})$&        &   &       O  &X&  \\
$S(B\to K_S \pi^0 \gamma)$                    &        &   &          & &X \\
$\beta$                                  &        &   &X-$CKM$   & & O \\
\hline
  \end{tabular}
\end{table}
Table~\ref{tab:golden_reach} reports the comparison of the experimental sensitivity today and with 75ab$^{-1}$, showing that in most cases Super$B$ is able to measure the observables with a few percent accuracy. 
As already mentioned in the previous section, these decays are very difficult or impossible to reconstruct at the LHC. Even in the clean environment of Super$B$ the selection is experimentally challenging and to suppress backgrounds to an acceptable level the recoil technique is often necessary, in which the other $B$ in the $B\bar B$ event is reconstructed in either a semileptonic or hadronic decay~\cite{kstnunu}.

\begin{table}[!h]
\begin{tabular}{ccc}
\hline
Mode & \multicolumn{2}{c}{Sensitivity} \\
   & Current & Expected ($75 \ {\rm ab}^{-1}$)  \\
\hline
 ${\cal B}(B \to X_s \gamma)$ & 7\% & 3\% \\
 $A_{CP}(B \to X_s \gamma)$ & 0.037 & 0.004--0.005 \\
 ${\cal B}(B^+ \to \tau^+ \nu)$ & 30\% & 3--4\% \\
 ${\cal B}(B^+ \to \mu^+ \nu)$ & not measured & 5--6\% \\
 ${\cal B}(B \to X_s l^+l^-)$ & 23\% & 4--6\% \\
 $A_{{\rm FB}}(B \to X_s l^+l^-)_{s_0}$ & not measured & 4--6\% \\
 ${\cal B}(B \to K \nu \overline{\nu})$ & not measured & 16--20\% \\
 $S(B\to K^0_S \pi^0 \gamma)$ & 0.24 & 0.02--0.03 \\
 \hline
 \end{tabular}
 \caption{
\label{tab:golden_reach}
 Comparison of current experimental sensitivities with those expected
 at Super$B$  ($75 \ {\rm  ab}^{-1}$).
 Only a small selection of observables is shown.
 Quoted sensitivities are relative uncertainties if given as a percentage,
 and absolute uncertainties otherwise.
 For more details, see Refs.~\cite{cdr,Browder:2007gg,Browder:2008em}.
    \label{tab:ukekb-comp}
  }
\end{table}

\begin{figure}[!h]
  \begin{center}
    \includegraphics[width=4.cm,angle=-90]{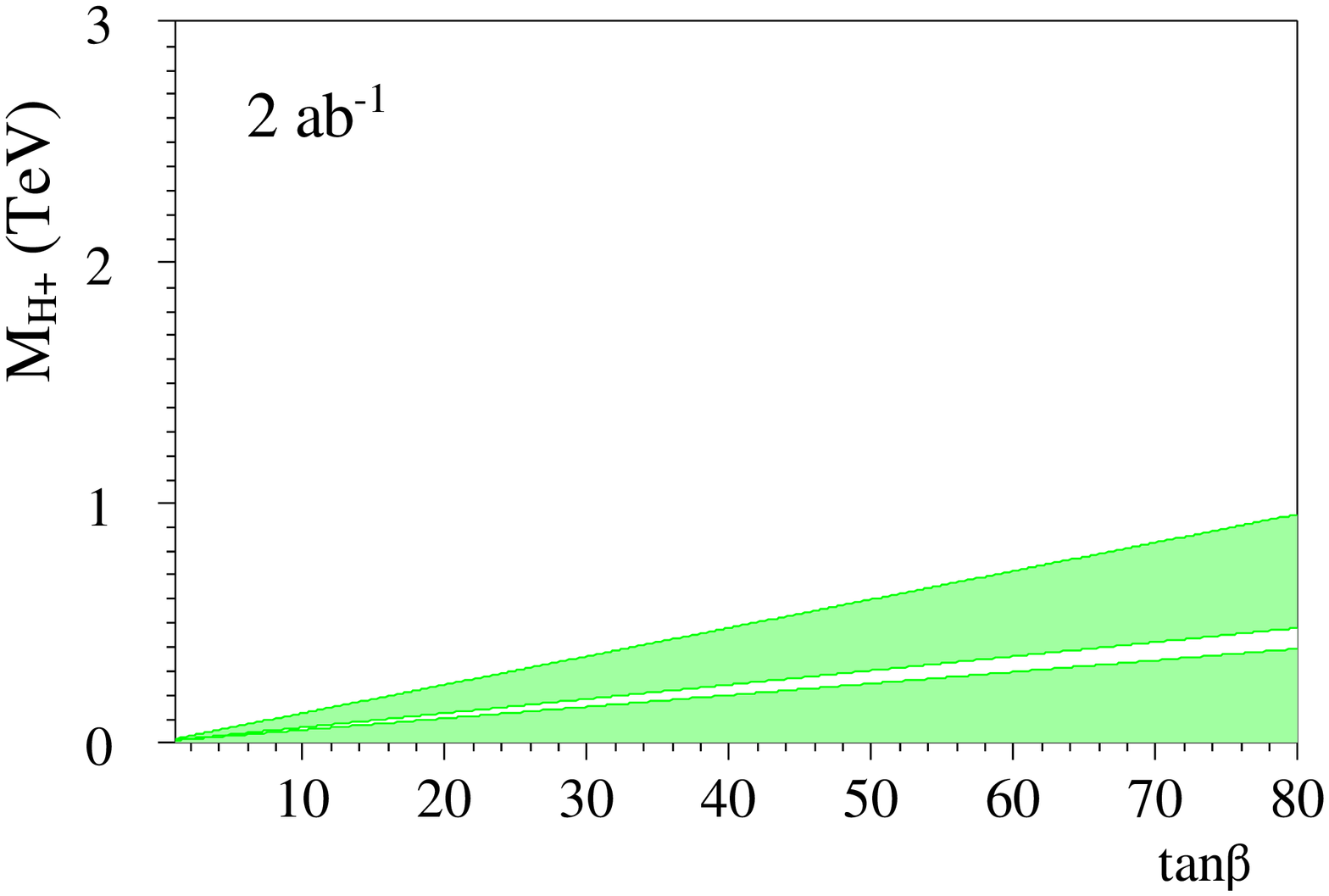}
    \includegraphics[width=4.cm,angle=-90]{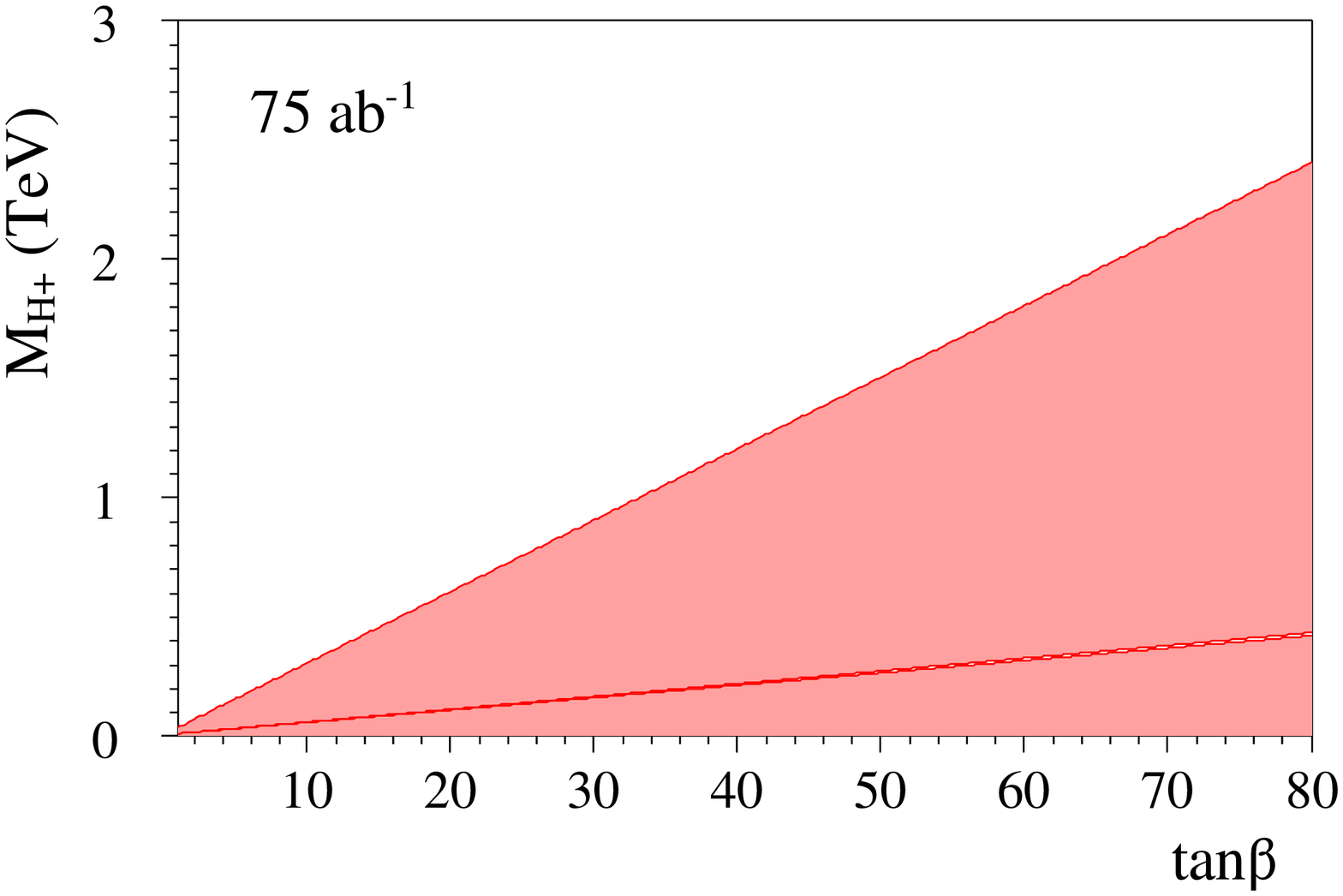}
    \caption{\label{fig:btaunu}
      Exclusion regions in the $m(H^+)$--$\tan\beta$ plane arising from the
      combinations of the measurement of ${\cal B}(B \to \tau \nu)$ and
      ${\cal B}(B \to \mu \nu)$ using 2 ab$^{-1}$ (left) and 75 ab$^{-1}$ (right). It is assumed that the experimental result is consistent with the Standard Model.
    }
  \end{center}
\end{figure}

As an example of process sensitive to NP we consider the decays $B\rightarrow l\nu$ with $l=\tau,\mu$, whose rates are strongly affected by a charged Higgs in a scenario with large $\tan\beta$. For example, in the two Higgs doublet model the effect of the charged Higgs is a rescaling of the SM branching fraction by a factor $(1-\tan^2\beta(M^2_B/M^2_{H^+}))^2$~\cite{hou}, where $M_{H^+}$ is the mass of the Higgs boson.
Fig.~\ref{fig:btaunu} gives the exclusion regions in the $M_{H^+}-\tan\beta$ plane from a measurement of $\mathcal{B}(B\rightarrow l\nu)$ with 2ab$^{-1}$ and 75ab$^{-1}$, assuming that the result is centered at the SM prediction. 
In scenarios with large $\tan\beta$, for example $\tan\beta\sim 50$, Super$B$ can push the lower bound on $M_{H^+}$ from the hundreds of GeV region up to almost 2 TeV.

\vspace{0.2cm}

\subsection{Time-dependent $CP$ asymmetry in penguin-dominated modes}
New Physics can be probed in mixing-induced $CP$ violation of processes dominated by $b\rightarrow s$ penguin loops. In the SM the time dependent $CP$ asymmetry of these decays should measure $\sin 2\beta$\footnote{Modulo a sign depending on the $CP$ content of the final state.} up to small corrections, i.e. $\Delta S\equiv\sin 2\beta|_{b\rightarrow s}-\sin 2\beta\simeq 0$. However, NP particles in the loops can cause measurable deviations from the SM prediction. The potential of this approach depends on the accuracy of the SM prediction for $\Delta S$ and on the experimental uncertainty on $\sin 2\beta|_{b\to s}$ for the individual channels. The decays where the expected deviation $\Delta S$ and the associated theoretical uncertainty are the smallest are $\eta' K^0$, $\phi K^0$ and $K_SK_SK_S$, making them the theoretically cleanest probes of NP. At present it appears that the reconstruction of these modes at a hadronic machine is at least challenging~\cite{Browder:2008em}.

The current experimental errors are still large compared to the theoretical uncertainties and no significant deviations from the SM predictions are observed. By extrapolating the existing measurements of a wide range of channels, one concludes that a data sample of at least 50 ab$^{-1}$ is necessary to reduce the experimental errors on $\Delta S$ to the level of 0.01-0.03, close to the current theory precisions of the cleanest modes. Therefore at a super$B$-factory these processes will be sensitive probes of NP.

\subsection{Precise measurement of the CKM parameters}
\begin{figure}[h!]
\begin{center}
\includegraphics[width=0.45\textwidth]{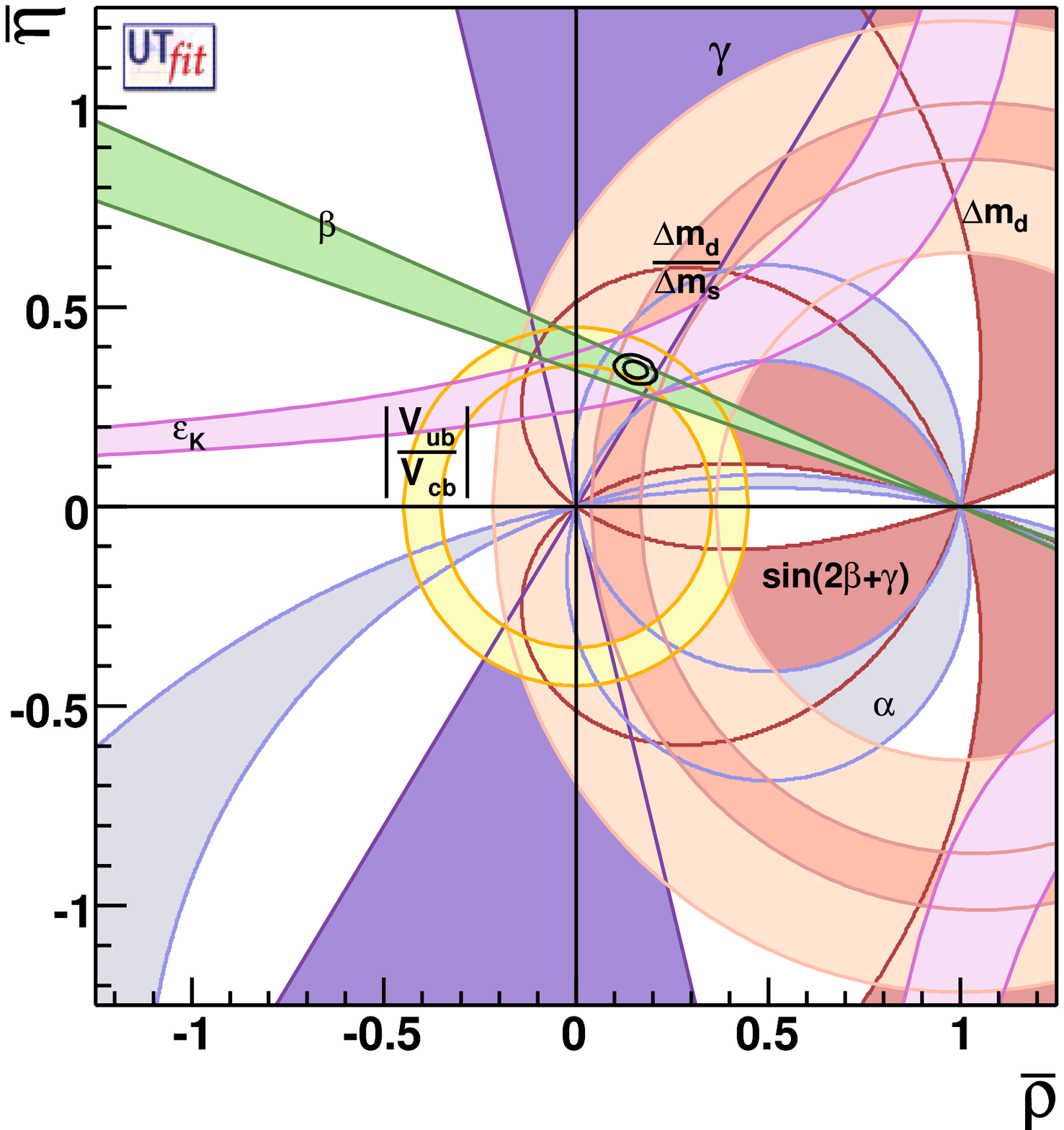}
\includegraphics[width=0.45\textwidth]{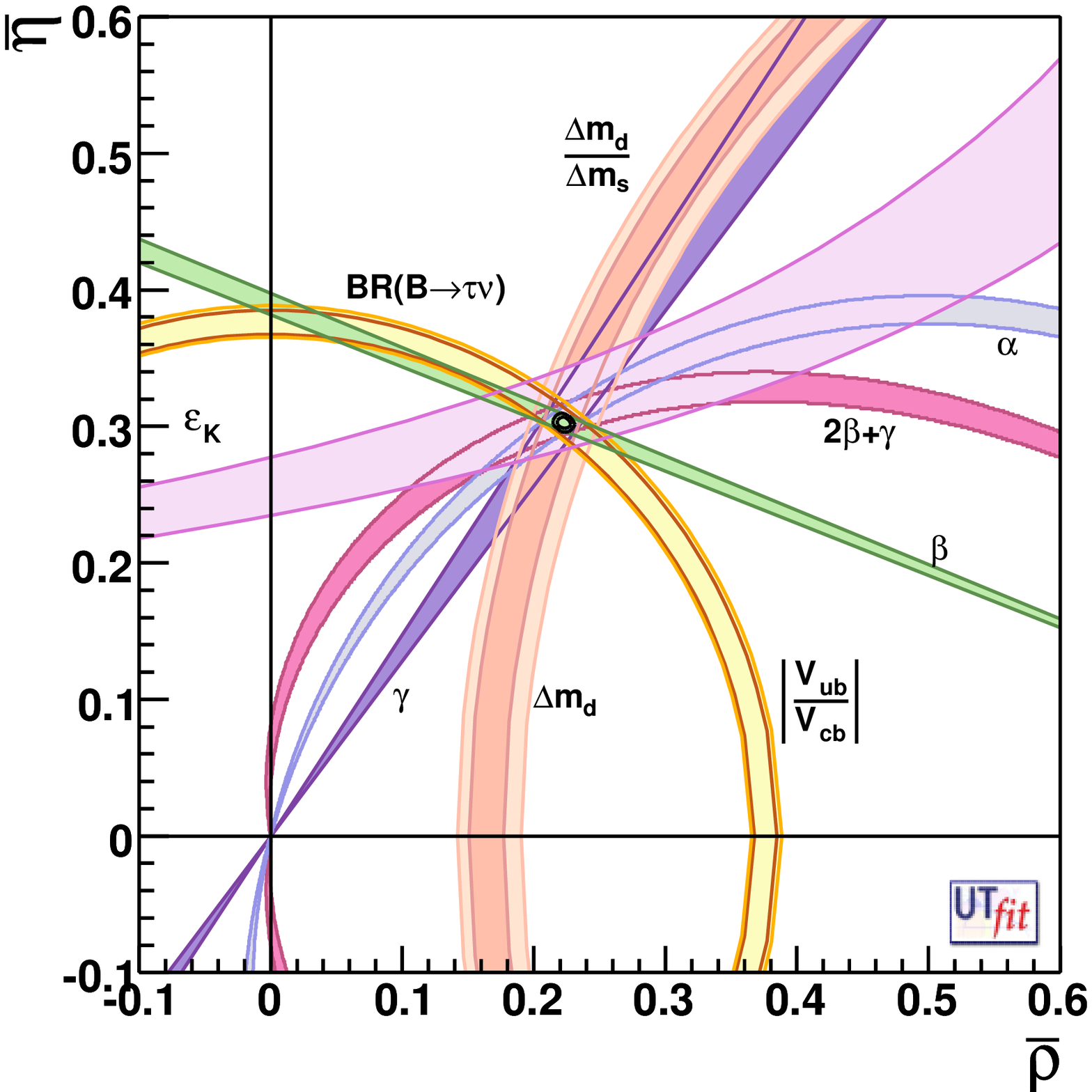}
\caption{
  Allowed regions corresponding to 95\% probability for $\bar{\rho}$ and $\bar{\eta}$ selected by different constraints, assuming present central values and errors (left) or a scenario with perfect agreement among the measurements and with the errors expected at a Super$B$ (right). }
\label{fig:rhoeta_cons}
\end{center}
\end{figure}

Super$B$ can dramatically improve our knowledge of the CKM matrix thanks to the possibility of performing a wide range of measurements which constrain its elements. 
Table~\ref{tab:ckm_precision} compares the errors of the CKM parameters obtained from the SM fit using the experimental and theoretical information available today and at a Super$B$, showing that the uncertainties would be reduced by a factor 10. The current constraints in the $\bar{\rho}-\bar{\eta}$ plane are reported in the left plot of fig.~\ref{fig:rhoeta_cons}, while the right plot shows the impressive improvement expected with a dataset of 50ab$^{-1}$ in a scenario where there is perfect agreement with the SM predictions. A precise knowledge of the CKM matrix is important per se, but it is also a powerful tool to spot inconsistencies in the SM and evidence of NP. Several measurements used for the determination of $\bar{\rho},\bar{\eta}$ can in fact be affected by the presence of NP, revealing itself as an inconsistency in the $\bar\rho-\bar\eta$ plane. 

\begin{table}[h!]
  \caption{
    Uncertainties of the CKM parameters obtained from the Standard Model fit
    using the experimental and theoretical information available today (left)
    and at the time of Super$B$\ with a dataset of 50ab$^{-1}$ (right).
  }
  \begin{center}
    \begin{tabular}{lll}
      \hline
      Parameter               &   SM Fit today        &  SM Fit at Super$B$ \\
      \hline
      $\overline {\rho}$      &   $0.163\pm 0.028$    &  $\pm 0.0028$      \\
      $\overline {\eta}$      &   $0.344\pm 0.016$    &  $\pm 0.0024$      \\
      $\alpha$ ($^{\circ}$)   &   $92.7\pm4.2$        &  $\pm 0.45$        \\
      $\beta$ ($^{\circ}$)    &   $22.2\pm0.9$        &  $\pm 0.17$        \\
      $\gamma$ ($^{\circ}$)   &   $64.6\pm4.2$        &  $\pm 0.38$        \\
      \hline
    \end{tabular}
  \end{center}
  \label{tab:ckm_precision}
\end{table}

\subsection{Lepton flavor violation in $\tau$ decays}
Super$B$ is able to explore a significant portion of the parameter space in many NP scenarios by searching for lepton flavor violation (LFV) in transitions between the third and first or second lepton generations, complementing studies in the muon sector such as the search for  $\mu\rightarrow e\gamma$ being performed by the MEG experiment~\cite{meg}.
Compared to the potential of the current $B$-factories considered together, with a data sample of 75ab$^{-1}$ Super$B$ can increase the sensitivity by more than a factor 7 in the worst hypothesis of background-dominated analyses even assuming no improvement in the analysis techniques. For analyses which are background-free the sensitivity is at least 50 times better. Moreover, the baseline Super$B$ machine design incorporates the polarization of the electron beam (up to 85\%), which will produce polarized $\tau$ leptons. The polarization can be exploited either to improve the background-signal separation, or to better determine the features of the LFV interaction once it is observed. Table~\ref{tab:tau} summarizes the sensitivities for various LFV decays together with the current world average upper limits~\cite{pdg09}. In a number of NP models the branching ratios of flavor-violating $\tau$ decays can be enhanced up to $\mathcal{O}(10^{-8})$, just below the current $B$-factories reach and beyond the possibility of LHC~\cite{tau_lhc} but well above the reach of Super$B$. Other sensitive probes of NP include tests of lepton flavor universality and the search of $CP$ violation in $\tau$ decays.

\begin{table}[ht]
\caption{Expected 90\% CL upper limits on representative LFV $\tau$ lepton decays with 75ab$^{-1}$ and current world average upper limits. }
\begin{center}
\begin{tabular}{lcc}
\hline
Process           &Sensitivity at Super$B$&Current limit\\
\hline
$\mathcal{B}(\tau\rightarrow\mu\gamma)$  & $2\times 10^{-9}$ &  $4.5\times 10^{-8}$\\
$\mathcal{B}(\tau\rightarrow e\gamma)$  & $2\times 10^{-9}$ &  $1.1\times 10^{-7}$\\
$\mathcal{B}(\tau\rightarrow\mu\mu\mu)$  & $2\times 10^{-10}$ &  $3.2\times 10^{-8}$\\
$\mathcal{B}(\tau\rightarrow eee)$  & $2\times 10^{-10}$ &  $3.6\times 10^{-8}$\\
$\mathcal{B}(\tau\rightarrow\mu\eta)$  & $4\times 10^{-10}$ &  $6.5\times 10^{-8}$\\
$\mathcal{B}(\tau\rightarrow e\eta)$  & $6\times 10^{-10}$ &  $9.2\times 10^{-8}$\\
$\mathcal{B}(\tau\rightarrow eK^0_S)$  & $2\times 10^{-10}$ &  $3.3 \times 10^{-8}$\\
$\mathcal{B}(\tau\rightarrow\mu K^0_S)$  & $2\times 10^{-10}$ &  $4.0\times 10^{-8}$\\
\hline
\end{tabular}
\end{center}
\label{tab:tau}
\end{table}

\subsection{Charm physics}
Super$B$ can operate as a charm factory at the CM energy of both the $\Upsilon(4S)$ and the $\Psi(3770)$, where the quantum correlations in the coherent production of $D^0\bar{D^0}$ can be exploited~\cite{valencia}. 
The charm production cross section at the $\Upsilon(4S)$ CM energy is comparable to the $B\bar B$ cross section, $\sigma(e^+e^-\to c\bar c)\sim 1.3$~nb. At production threshold the luminosity is smaller by a factor of ten with respect to the baseline value, but it is partially compensated by a cross section about three times larger. 
The recent observation of the $D^0-\bar{D^0}$ mixing opens a unique window to the search of $CP$ violation in the charm sector, whose observation would provide a strong hint of physics beyond the SM. The program also includes the search of $CP$ violation in time-independent measurements and the study of rare and forbidden charm decays, as well as precise measurements of CKM matrix parameters~\cite{cdr,valencia}.

\section{The detector}
To reach the required luminosity of $10^{36}$~cm$^{-2}$s$^{-1}$ Super$B$ exploits a new collision scheme which is based on a small collision area, very small $\beta^*_y$ at the interaction point, large Piwinsky angle and the 'crab waist' scheme~\cite{cdr,raimondi,raimondi2}. This novel approach has several advantages, most notably the fact that the very large improvement in luminosity is achieved with beam currents and wall plug power similar to those of the current $B$-factories, and with limited background rates. In the current layout the accelerator consists of 4~GeV/7~GeV positron/electron beams, corresponding to a CM boost $\beta\gamma\sim 0.28$ in the lab frame (half the value in BaBar).

The Super$B$ detector concept is based on the BaBar design~\cite{babardet} with some modifications required to deal with the reduced boost and higher event rates.
A number of components of the SLAC $B$-factory can be reused, resulting in a significant reduction of costs. This includes parts of the PEP-II accelerator complex, the super-conducting solenoid, the CsI(Tl) crystals of the barrel electromagnetic calorimeter (EMC) and the quartz bars of the Cherenkov particle identification system (DIRC). 

In the remainder of this section a short description of the detector under development is provided, focussing on a few aspects where Super$B$ differs from BaBar. 
 
The tracking system is composed of a silicon vertex detector (SVT) surrounded by the drift chamber (DCH). To maintain sufficient $\Delta t$ resolution for time-dependent $CP$ violation measurements with the reduced Super$B$ boost, the vertex resolution is improved by reducing the radius of the innermost layer of SVT (layer-0) down to about 1.5~cm, which is the lower limit allowed by the background rates according to preliminary simulation studies. Two main options are being considered for the layer-0, CMOS monolithic active pixels thin sensors~\cite{rizzo} or hybrid pixels detectors~\cite{alice}, while the outer silicon layers are made of microstrips silicon sensors.
The starting point for the DCH layout is the BaBar drift chamber, though the design optimization process may eventually end up in a quite different device. Anticipated improvements include a lighter, carbon-fiber, mechanical structure and faster readout electronics. 
The hadron particle identification system is placed just outside the DCH and will make use of the radiator quartz bars of the BaBar DIRC, 
with the old PMTs replaced by fast pixelated PMTs and the imaging region reduced in size to control the background rates. 
The EMC can reuse the barrel portion of the BaBar EMC, made of 5760 CsI(Tl) crystals. This fact is very important because the barrel EMC is the most expensive element of the detector. In contrast, the CsI(Tl) crystals of the forward endcap will be replaced with L(Y)SO crystals, which are suitable for their excellent radiation hardness, fast decay time, small Moli\`ere radius and relatively high light yield. The outermost detector system is the Instrumented Flux Return for the detection of muons and neutral hadrons. The Resistive Plate Chambers and Limited Streamer Tubes used in BaBar will be replaced by extruded scintillator bars \`a la MINOS~\cite{minos} and the amount and distribution of the absorber (iron or brass) will be optimized.

Two additional systems are being considered to possibly improve the performance of the detector: a forward particle identification device placed between the DCH and the forward endcap EMC, and a backward EMC calorimeter. Two candidates for the PID system are currently being compared, an aerogel radiator RICH and a time-of-flight system using a sheet of fused silica radiator~\cite{cdr}. The main purpose of the backward EMC would be to reject background events by detecting extra energy in that region, and therefore the energy resolution is not a critical parameter. A relatively simple device made of lead plates and scintillating strips may be adequate.

A Technical Design Report (TDR) of the project is in preparation and is expected to be completed within two years. 

\section{Summary}
Super$B$ is a next generation asymmetric energy $e^+e^-$ flavor factory operating mainly at the $\Upsilon(4S)$ CM energy but with the possibility to run in a wide range of energies, from the $\tau\tau$ threshold up to the $\Upsilon(5S)$. The main goal of the experiment is to search for evidence of physics beyond the Standard Model in the decays of heavy quarks and leptons, and to constrain its parameters. Super$B$ and the LHC are largely complementary in their effort to observe NP effects. In five years at the baseline luminosity of $10^{36}$~cm$^{-2}$s$^{-1}$ Super$B$ can collect 75ab$^{-1}$, which correspond to about $8\times 10^{10}$ $B\bar B$ pairs, $5\times 10^{10}$  $\tau$ lepton pairs and $1\times 10^{11}$ $c\bar c$ pairs. This dataset allows to explore and test the flavor sector of the Standard Model with unprecedented precision. The project has entered the TDR phase which is expected to last two years.

\section{Acknowledgments}
The author would like to thank the conference organizers and the Super$B$ community for the opportunity to give this talk.

\end{document}